\documentclass[prb,aps,twocolumn,showpacs, floatfix,preprintnumbers]{revtex4}

\usepackage[applemac]{inputenc}
\usepackage[T1]{fontenc} 
 \usepackage{amsmath} 
\usepackage{subfigure}
\usepackage{lipsum}
\usepackage{amsfonts} 
\usepackage{amssymb, mathrsfs}
\usepackage{braket}
\usepackage{graphicx} 
\usepackage[usenames]{color} 

\usepackage{bbm}

\def\beq{\begin{equation}}
\def\eeq{\end{equation}}
\def\bsp{\begin{split}}
\def\esp{\end{split}}
\def\bea{\begin{eqnarray}}
\def\eea{\end{eqnarray}}
\def\ba{\begin{array}}
\def\ea{\end{array}}

\def\lb{\left(}
\def\rb{\right)}

\def\l.{\left.}
\def\r.{\right.}

\def\part{\partial}

\def\bra#1{{\cal{h}} #1 \mid}
\def\ket#1{\mid #1 {\cal{i}}}

\makeatletter

\newcommand{\Rmnum}[1]{\expandafter\@slowromancap\romannumeral #1@}
\makeatother
\begin{document}
\title{Rotating Entangled States of an Exchange-Coupled Dimer of Single-Molecule Magnets.}
\author{S. A Owerre}
\email{solomon.akaraka.owerre@umontreal.ca}
\affiliation{Groupe de physique des particules, D\'epartement de physique,
Universit\'e de Montr\'eal,
C.P. 6128, succ. centre-ville, Montr\'eal, 
Qu\'ebec, Canada, H3C 3J7 }

\begin{abstract}
\section*{Abstract}  
An antiferromagnetically exchange-coupled  dimer of single molecule magnets which possesses a large spin tunneling has been investigated. For this system the ground and first excited states are entangled states and the Hamiltonian is effectively similar to that of a two-state system at $2s^{\text{th}}$ order in perturbation theory, thus this system can be mapped to an entangled pseudospin 1/2 particles. We study the effects of interaction and rotation of this system  about its staggered easy-axis direction . The corresponding Hamiltonian of a rotated two-state entangled spin system is derived with its exact low-energy eigenstates and eigenvalues. We briefly discuss the effect of a dissipative environment on this rotated two-state system.

\end{abstract}


\maketitle


{\it{Introduction}-}
Macroscopic quantum tunneling and coherence of a single, molecular, magnetic large spin system ( such as Mn$_{12}$ and Fe$_{8}$) have been the subject of interest for decades\cite{cg,ct}. These systems are composed of several molecular magnetic ions, whose spins are coupled by intermolecular interactions giving rise to  an effective single large spin. Their tunneling behaviour as well as quenching of tunneling have been studied extensively by spin coherent state path integral formalism\cite{kl,cg,dl,ct,vh} and experimental method\cite{af}. In its simplest form, the Hamiltonian is comprised of two terms, one term $\hat{H}_{\parallel}$, which commutes with the $z$-component of the spin and the other term $\hat{H}_{\perp}$, which does not commute with the $z$-component of the spin is responsible for the tunneling splitting between the two degenerate ground states $\ket{\pm s}$ of $\hat{H}_{\parallel}$. Due to tunneling, the ground and the first-exited states become the symmetric and antisymmetric linear superpositions of $\ket{\pm s}$. In recent years, the problem of a molecular magnet which is free to rotate about the easy axis has attracted considerable attention. This problem involves the conservation of the total angular momentum  due the fact that the rotating nanomagnet couples to its mechanical motion in an equal and opposite directions. It has been studied experimentally for free magnetic clusters \cite{solo1, wznw} and magnetic microresonators \cite{akason, solo2}. A theoretical study has also been investigated \cite{cl,cl1}, which shows that the coupling of the mechanical motion and the spin renormalizes the magnetic anisotropy and increases the tunneling splitting.

 The tunneling phenomenon of nanomagnet is not restricted to single molecule magnets (SMMs). In many cases of physical interest, interactions between two large spins are taken into account. These interactions can be either ferromagnetic, which aligns the neighbouring spins or antiferromagnetic, which anti-aligns the neighbouring spins. One physical system in which these interactions occur is the dimerized molecular magnet [Mn$_4$]$_2$. This system comprises two Mn$_4$ SMMs of equal spins $s_1=s_2=9/2$, which are coupled antiferromagnetically. The phenomenon of quantum tunneling of spins in this system has been be studied both numerically and experimentally\cite{akason1,aff}. A theoretical study via spin coherent state path integral formalism has been reported recently \cite{ams} and perturbation theory has also been investigated\cite{kim2, bab1}. In this case the situation is quite different from that of SMMs in that quantum tunneling is achieved via entangled antiferromagnetic states.
For a free antiferromagnetically exchange-coupled dimer, it is interesting to ask how does the spins couple with the mechanical motion of the system, what is the effective two-state system and what is the effect of a dissipative environment on such rotating particles. These questions are yet an unsolved problem. We will try to address them in this paper.

{\it{Tunneling of antiferromagnetically exchange-coupled dimer, two-state system and dissipative environment }-}
In this section we will briefly review the model of antiferromagnetically exchange-coupled  dimer of SMMs with large equal spins, and its tunneling effect.  For this system, the simplest form of the Hamiltonian in the absence of an external magnetic field can be written as
\begin{align}
\hat{H}&= -\mathcal{D}(\mathcal{\hat{S}}_{1,z}^2+\mathcal{\hat{S}}_{2,z}^2)  + \mathcal{J}\mathcal{\hat{\boldsymbol{S}}}_{1}\cdot\mathcal{\hat{\boldsymbol{S}}}_{2} = \hat{H}_0 + \hat{H}_{int}
\label{1}\end{align}
where $\mathcal{J}>0$ is the antiferromagnetic exchange coupling, $\mathcal{J}<0$ is the ferromagnetic exchange coupling and $\mathcal{D}\gg\mathcal{J}>0$ is the easy-axis anisotropy constant, $\mathcal{S}_{i,z}, i=1,2$ is the projection of the component of the spin along the $z$ easy-axis.  It has been demonstrated by density-functional theory\cite{parkma} that this simple model can reproduce experimental results in  [Mn$_4$]$_2$ dimer with $\mathcal{D}=0.58 K$ and $\mathcal{J}=0.27 K$ . For spin 1/2, this model can be used to model a two-qubit of quantum dots interacting via a tunneling junction. It also plays a crucial role in quantum CNOT gates and SWAP gates \cite{dd}. The total $z$-component of the spins $\mathcal{S}_{z}=\mathcal{S}_{1,z}+\mathcal{S}_{2,z}$ is a conserved quantity, thus  the Hamiltonian is invariant under rotation about this direction. However, the individual $z$-component spins $\mathcal{S}_{1,z}, \mathcal{S}_{2,z}$ and the staggered configuration $\mathcal{S}_{1,z}-\mathcal{S}_{2,z}$ are not conserved.   In the absence of $\hat{H}_{int}$, the Hamiltonian is four-fold degenerate corresponding to the states where the individual spins are in their highest weight or lowest weight states, $|\hskip-1 mm\uparrow, \uparrow\rangle, |\hskip-1 mm\downarrow, \downarrow\rangle, |\hskip-1 mm\uparrow, \downarrow\rangle, |\hskip-1 mm\downarrow, \uparrow\rangle$, where $\ket{\uparrow, \downarrow}=\ket{\uparrow}\otimes\ket{\downarrow}\equiv \ket{s,-s}$ etc, with the exchange interaction term $\mathcal{J}$, the two ferromagnetic states $\ket{\uparrow, \uparrow}$ and $\ket{\downarrow, \downarrow}$  are still degenerate but the antiferromagnetic states  $\ket{\uparrow, \downarrow}$ and $\ket{\downarrow, \uparrow}$ are not. Perturbation theory \cite{kim2, bab1}, spin coherent state path integral formalism and effective potential mapping\cite{ams} showed that these two antiferromagnetic states are linked to each other at $2s^{\text{th}}$ order in $\hat{H}_{int}$, that is $\mathcal{J}^{2s}$. Thus, these two states reorganize into symmetric and antisymmetric linear superposition. The quantum spin Hamiltonian at $2s^{\text{th}}$ order  can then be written effectively as
\bea
\hat H \psi_{\pm}=\mathcal{E}_{\pm}\psi_{\pm}, \quad \Delta=\mathcal{E}_{+}-\mathcal{E}_{-} \sim \lb\frac{\mathcal{|J|}}{4\mathcal{D}}\rb^{2s}
\label{3.17}
\eea
where
\begin{align}
\psi_{-}&= \frac{1}{\sqrt{2}}\lb\ket{\uparrow, \downarrow}-\ket{\downarrow,\uparrow}\rb,\label{3.17a}\\\psi_{+}&= \frac{1}{\sqrt{2}}\lb\ket{\uparrow, \downarrow}+\ket{\downarrow,\uparrow}\rb\nonumber, 
\end{align}
\begin{figure}[ht]
\centering
\includegraphics[scale=0.35]{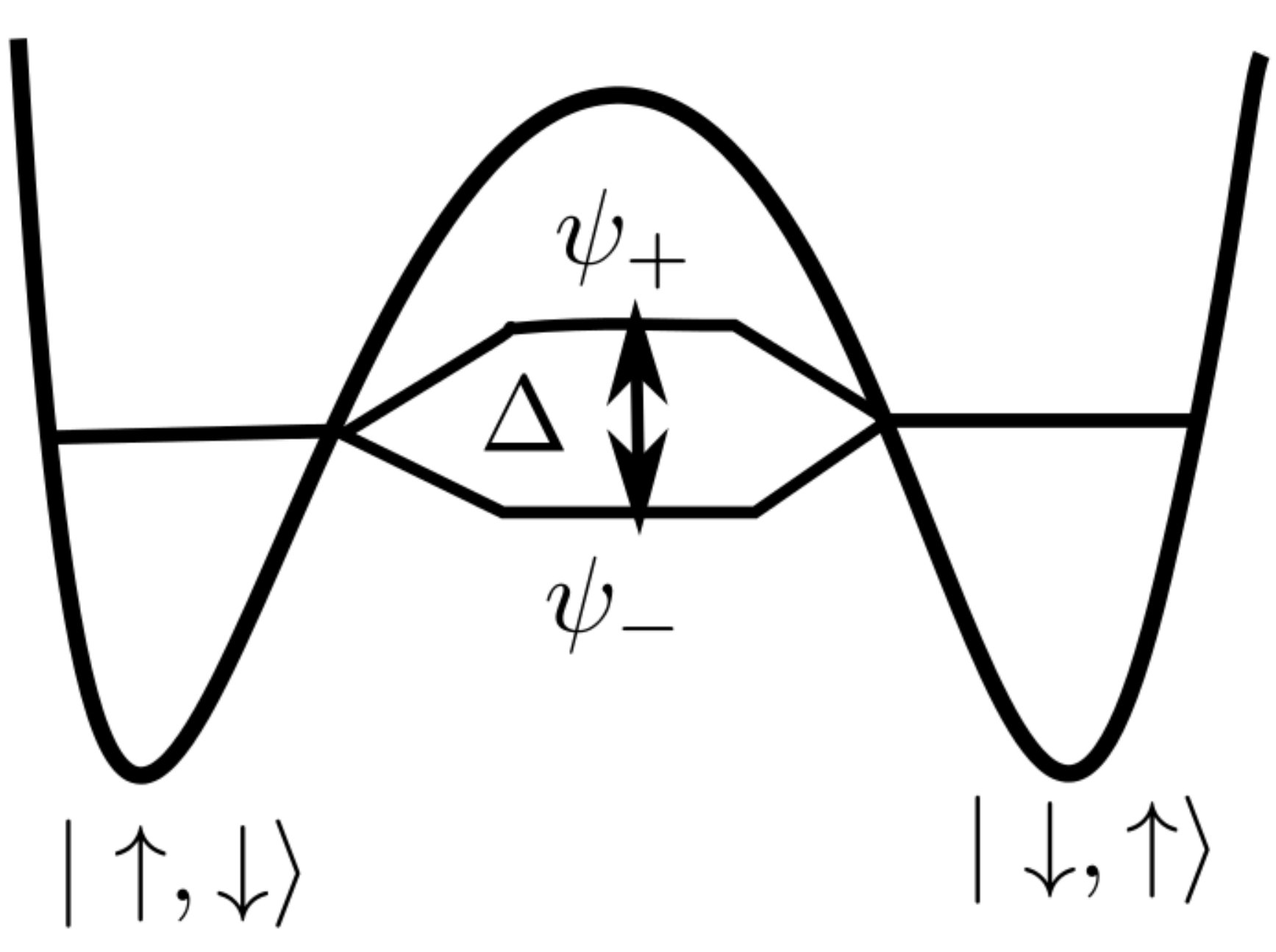}
\label{pont}
\caption{ Sketch of the potential energy with a barrier. The two antiferromagnetic states $\ket{\uparrow,\downarrow}$ and $\ket{\downarrow,\uparrow}$ are localized at the left and the right minimum of the potential, due to tunneling these states reorganize into antisymmteric and symmetric combination with an energy splitting separating them.}
\end{figure}

 The ground state corresponds to the maximally entangled antisymmetric combination while the first excited state corresponds to the maximally entangled symmetric combination for half-odd integer spins. For integer spins these roles are reversed. In quantum computing terminology for spin 1/2 particles, these energy eigenstates are entangled states which play a decisive role in quantum information processes, such as quantum teleportation and quantum register. There are equal probabilities of measuring either $\ket{\downarrow,\uparrow}$ or $\ket{\uparrow,\downarrow}$, that is 1/2. The eigenvalue equation, Eq.\eqref{3.17} is effectively similar to that of a two-state system, thus this system can be mapped to an entangled pseudospin 1/2 particles whose motion is restricted to the subspace of the total Hilbert space. A convenient  way of doing this mapping is by constructing the components of a matrix operator with the two-qubit states  $\ket{\downarrow,\uparrow}$ and $\ket{\uparrow,\downarrow}$ as
 \begin{align}
 \hat{\alpha}_x &= \ket{\uparrow,\downarrow}\bra{\downarrow,\uparrow}+\ket{\downarrow,\uparrow}\bra{\uparrow,\downarrow}   \nonumber\\
 \hat{\alpha}_y &= -i\ket{\uparrow,\downarrow}\bra{\downarrow,\uparrow}+i\ket{\downarrow,\uparrow}\bra{\uparrow,\downarrow}  \label{2.4}\\
 \hat{\alpha}_z &=  \ket{\uparrow,\downarrow}\bra{\uparrow,\downarrow}-\ket{\downarrow,\uparrow}\bra{\downarrow,\uparrow}  \nonumber\\
\hat{I} &= \ket{\uparrow,\downarrow}\bra{\uparrow,\downarrow}+\ket{\downarrow,\uparrow}\bra{\downarrow,\uparrow}  \nonumber
 \end{align}
 
 These matrices are entangled since they cannot be separated as $\hat{\alpha}_{1,x} \otimes \hat{\alpha}_{2,x}$, etc. It is noted that in the two-qubit form they satisfy the usual commutation and anti-commutation relations $[\hat{\alpha}_{k},\hat{\alpha}_{j}]=2i\epsilon_{ijk}\hat{\alpha}_{k}$ and $\lbrace\hat{\alpha}_{k},\hat{\alpha}_{j}\rbrace=2\delta_{ij}$ .
In the matrix representation they are $4\times4$ sparse matrices which are not Pauli matrices but their subspace contains the usual  $2\times 2$ Pauli matrices. The Hamiltonian $\hat{H}$ can now be projected unto the two qubit states $\ket{\downarrow,\uparrow}$ and $\ket{\uparrow,\downarrow}$:
\bea
\hat{H}_{\alpha} = \sum_{m,n=\pm s}\ket{m,-m}\hat{H}_{mn}\bra{n,-n}
\eea
where $\hat{H}_{mn}=\braket{m,-m|\hat{H}|n,-n}$.
Using Eqns.\eqref{3.17} and \eqref{3.17a}, the matrix elements are found to be
\begin{align}
\braket{\downarrow,\uparrow|\hat{H}|\downarrow,\uparrow}&= \braket{\uparrow,\downarrow|\hat{H}|\uparrow,\downarrow} = 0\nonumber\\
\braket{\downarrow,\uparrow|\hat{H}|\uparrow,\downarrow}&= \braket{\uparrow,\downarrow|\hat{H}|\downarrow,\uparrow} = \Delta/2
\end{align}
Thus the projected Hamiltonian becomes
 \beq
\hat{H}_{\alpha}  =\frac{\Delta}{2} \hat{\alpha}_x
\label{fre}
\eeq
with its eigenvalues given by $\lbrace-\Delta/2,\Delta/2\rbrace$. 
In many cases of physical interest, the spin interacts with its environment which is usually modeled as a bath of bosons. The environmental effect on the Hamiltonian, Eq.\eqref{fre} is written in the usual form\cite{chud1}.
\begin{align}
\hat{H}= \frac{\Delta}{2} \hat{\alpha}_x + I\sum_{k}\epsilon_k \hat{b}_{k}^{\dagger} \hat{b}_k+\chi S_z
= \hat{H}_\alpha+\hat{H}_B  + H_{int}
\end{align}
where $\hat{H}_\alpha=\Delta\hat{\alpha}_x/2$,  $\hat{H}_B= \sum_{k}\epsilon_k \hat{b}_{k}^{\dagger} \hat{b}_k$,
 $H_{int}=\hat{\chi}\hat{S}_z$,  $\hat \chi=\sum_{k}\gamma_k (\hat{b}_k+\hat{b}_{k}^{\dagger})$, $\hat{S}_z=\hat{\alpha}_z/{2}$ and $\hat{b}_{k}^{\dagger},  \hat{b}_k$ are the
annihilation and creation operators of phonons with the wave number $k$.
In terms of the basis states, this Hamiltonian can be written as
\begin{align}
\hat{H}= (\ket{\uparrow,\downarrow}\frac{\Delta}{2}\bra{\downarrow,\uparrow} +h.c) +\ket{\downarrow,\uparrow}\hat{K}_{-}\bra{\downarrow,\uparrow} +\ket{\uparrow,\downarrow}\hat{K}_{+}\bra{\uparrow,\downarrow}
\end{align}
where 
$
\hat{K}_{\pm}= \sum_{k}\epsilon_k \hat{b}_{k}^{\dagger} \hat{b}_k\pm \hat\chi.
$
In the interaction picture we can solve for the time-evolution of $\hat{\alpha}_z(t)$ using the Heisenberg equation of motion with respect to $\hat{H}_\alpha $. The resulting expression is given by
\bea
\hat{\alpha}_z(t)= \hat{\alpha}_z\cos(\Delta t)+\hat{\alpha}_y\sin(\Delta t)
\eea

The coefficients of the trigonometric function are determined from the initial conditions at $t=0$.
The operator from which all other observables can be calculated is the reduced density operator. It is given by
\begin{align}
\hat{\rho}(t)=Tr_{B}\lb e^{-i\hat{H}t}\hat{\rho}(0)e^{i\hat{H}t}\rb
\label{aka6}
\end{align}
where $\hat\rho(0)=\hat{\rho}_{\alpha}\otimes \hat{\rho}_B$, where $\hat{\rho}_{\alpha}$ acts on the spin space and $\hat{\rho}_{B}=e^{-\beta\hat{H}_B}/{Z_B}$ is the density matrix of a free boson.
This system is exactly solvable in the limit $\Delta\rightarrow 0$ (independent boson model). In this limit the bosonic bath can be traced out in Eq.\eqref{aka6}. The object of interest in this model is usually the average values of the time evolution of the observables $\hat{\alpha}_i$, $i=x,y,z$. Using the interacting picture formulation of quantum mechanics, we have
\bea
\braket{\hat{S}_{+}(t)}= Tr_\alpha\lb\hat{U}^{\prime}(t,0)\rho(0)\hat{U}^{\prime}(0,t)S_{+}(0)\rb
\eea
where $\hat{S}_{+}=\hat{S}_x+i\hat{S}_y$, $\hat{U}^{\prime}(t,0)=Te^{-i\int_{0}^{t}dt\hat{H}^{\prime}(t)}$ and $\hat{H}^{\prime}(t)= e^{i\hat{H}_Bt}\hat{H}_{int}e^{-i\hat{H}_Bt}$.
A straight forward calculation of the trace yields
\bea
\braket{\hat{S}_{+}(t)}= C(t)\braket{\hat{S}_{+}(0)}, 
\eea
with $C(t)=\braket{Te^{-i\int_{0}^{t}\hat{\chi}^{\prime}(t)dt}}_0$ is the coherence factor between the states $\ket{\downarrow,\uparrow}$ and $\ket{\uparrow,\downarrow}$. By symmetry consideration we have 
$
\braket{\hat{S}_{-}(t)}= C(t)\braket{\hat{S}_{-}}(0),
$
hence
\bea
\braket{\hat{\alpha}_{x}(t)}= C(t)\braket{\hat{\alpha}_{x}(0)}
\eea

The evaluation of the equilibrium expectation value  with the use of Wicks theorem\cite{solo4} gives 
\bea
C(t)= e^{-\mathcal{D}(t)}, \quad \text{and}\thinspace |C(t)|= e^{-Re[\mathcal{D}(t)]}
\eea
\begin{align}
\mathcal{D}(t)&= \sum_k \frac{\gamma_k ^2}{\epsilon_k ^2}[(1+n_B)(1-e^{-i\epsilon_k t}) + n_B(1-e^{-i\epsilon_k t})\nonumber\\&-i\epsilon_k t], \quad n_B= (e^{\beta\epsilon_k}-1)^{-1}
\end{align}
\begin{figure}[ht]
\centering
\includegraphics[scale=0.35]{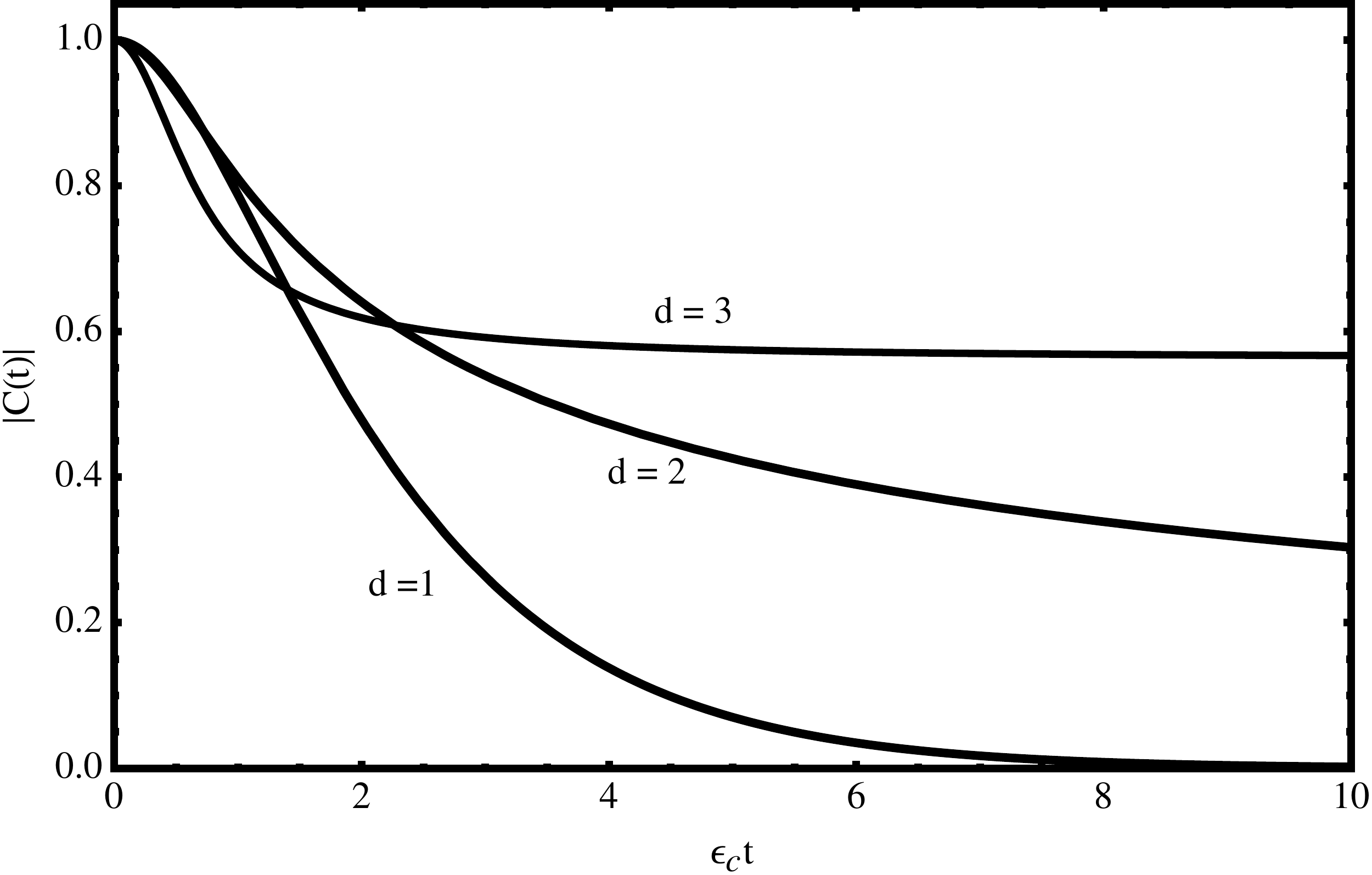}
\caption{ The plot of $|C(t)|$ against time for ohmic $d=1$ and super-hhimc $d=2,3$ dissipations. The function reaches a maximum of one and decays to zero at long times for the ohmic dissipation but it is never decays zero for the super-ohmic dissipation. }
\label{coh}
\end{figure}
In the continuum  limit we have 
\begin{align}
Re[\mathcal{D}(t)] = 2\int_0^{\infty} d\epsilon \frac{J(\epsilon)}{\epsilon ^2}\sin^2\lb\frac{\epsilon t}{2}\rb\cot\lb\frac{\beta\epsilon }{2}\rb
\label{aka3}
\end{align}
where the spectral density function $J(\epsilon)=N_d(\epsilon)\gamma(\epsilon)^2=\mathcal{K}_d\gamma_0^2\epsilon^de^{-\epsilon/\epsilon_c}$ and $N_d(\epsilon)$ is the $d$-dimensional density of phonon modes. For $d= 1,2, 3$ we have $\mathcal{K}_1=L/\pi c_s$, $\mathcal{K}_2=A/2\pi c_s^2$ and $\mathcal{K}_3=V/2\pi^2 c_s^3$, where $L,A, V$ are the length, area and volume of the system respectively, $c_s$ is the speed of sound. In Fig.\eqref{coh} we have plotted $|C(t)|$ for $d=1,2,3$, with $\beta\epsilon_c=1$. In the super-ohmic dissipation, $d= 2,3$, the coherence factor never decays to zero while for the ohmic dissipation $d=1$, the coherence factor completely decays to zero for large times $\epsilon_c t\gg1$, this is quite obvious from Eq.\eqref{aka3}. For $\Delta\neq0$, several techniques have been developed to study this model. The most elaborate functional integral analysis can be found in Ref.[20].

Suppose we apply a time varying staggered magnetic field along the easy $z$ axis on the exchange-coupled dimer model, the Zeeman Hamiltonian is of the form $\hat{H}_Z= -h(t)(\mathcal{\hat{S}}_{1,z}-\mathcal{\hat{S}}_{2,z})=-g\mu_Bh(\mathcal{\hat{S}}_{1,z}-\mathcal{\hat{S}}_{2,z})\cos\omega t$ , then projecting unto the two-qubit states, Eq.\eqref{fre} becomes
\bea
\hat{H}_{\alpha}  =\frac{\Delta}{2} \hat{\alpha}_x-2sg\mu_Bh\hat{\alpha}_z\cos\omega t
\eea
Transforming along the $y$-axis with the unitary operator  
\begin{align}
\hat{U}(\phi)=\exp(-i\phi\hat{\alpha}_y/2), \quad 
\phi=-\pi/2
\label{unitary}
\end{align} 
one obtains
\bea
\hat H_\alpha\rightarrow \hat{U}^{-1}(\phi)\hat H_\alpha \hat{U}(\phi)= -\frac{\Delta}{2}\hat{\alpha}_z-sg\mu_Bh\hat{\alpha}_x\cos\omega t
\eea
The corresponding wave function of this system is given by
\bea
\ket{\psi(t)}=\mathcal{C}_{\uparrow,\downarrow}(t)e^{-i\mathcal{E}_{+}t}\ket{\uparrow,\downarrow}+\mathcal{C}_{\downarrow,\uparrow}(t)e^{-i\mathcal{E}_{-}t}\ket{\downarrow,\uparrow}
\eea
with $|\mathcal{C}_{\uparrow,\downarrow}(t)|^2+|\mathcal{C}_{\downarrow,\uparrow}(t)|^2=1$ and $\mathcal{E}_{\pm}=\pm\Delta/2$. 
In the rotating wave approximation\cite{solo5, solo6}, the coefficients are given by
\begin{align}
\mathcal{C}_{\uparrow,\downarrow}(t)&=e^{-i{\Delta^{\prime}}t/2}\bigg[\cos\lb\frac{\Omega t}{2}\rb+i\frac{\Delta^{\prime}}{\Omega}\sin\lb\frac{\Omega t}{2}\rb\bigg]\nonumber\\
\mathcal{C}_{\downarrow,\uparrow}(t)&=ie^{i{\Delta^{\prime}}t/2} \frac{\Omega_R}{\Omega}\sin\lb\frac{\Omega t}{2}\rb
\end{align}
where
\begin{align}
\Delta^{\prime}=\Delta-\omega, \quad \Omega= \sqrt{\Omega_R ^2+\Delta^{\prime 2}} \quad\text{and} \quad \Omega_R = sg\mu_B h
\end{align}
Using these results the expectation values of the observables $\alpha_i$ are
\begin{align}
\braket{\hat{\alpha}_{x}}_t& = \frac{\Omega_R\Delta^{\prime}}{\Omega^2}(1-\cos\lb{\Omega t}\rb)\\
\braket{\hat{\alpha}_{y}}_t&=-\frac{\Omega_R}{\Omega}\sin\lb{\Omega t}\rb \\
\braket{\hat{\alpha}_{z}}_t&=\frac{\Delta^{\prime 2}}{\Omega^2}+\frac{\Omega_R ^2}{\Omega^2}\cos\lb{\Omega t}\rb
\end{align}

{\it{Rotation, interaction and environment}-}
 Rotation of a nanomagnet about the easy-axis  by an angle introduces an additional coupling to the spin Hamiltonian due the the mechanical motion of the system about the axis of rotation\cite{cl,cl1,cl3}. This coupling involves the angular momentum vectors of the rotating molecules\cite{solo7}. In this section, we will study the effect of rotating our model Hamiltonian about its easy-axis. In our simple dimer model, the ground state has a total spin of $\mathcal{S}_z=\mathcal{S}_{1,z}+\mathcal{S}_{2,z}=0$. Thus, the total $z$-component of the two SMMS is a conserved quantity, which directly implies that any rotation about this axis will leave the Hamiltonian invariant. We must seek for a direction on the easy-axis that does not commute with the Hamiltonian. A nontrivial rotation of Eq.\eqref{1} about the easy $z$-axis can be achieved by
 \bea
 \hat{\tilde{H}} = e^{-i(\mathcal{S}_{1,z}-\mathcal{S}_{2,z})\phi}\hat{H}e^{i(\mathcal{S}_{1,z}-\mathcal{S}_{2,z})\phi}
 \label{aka}
 \eea
 where $\phi=\phi_1-\phi_2$ is the relative angle on this axis and
 $\mathcal{S}_{i,z}\ket{\downarrow, \uparrow} \cong (-1)^{i}s\ket{\downarrow, \uparrow}$. This transformation can be physically realized in a spin-torque nano-oscillator with a two-state macroscopic spin (nanomagnet) which is free to rotate about its staggered easy-axis. A good example is that of  a spin-torque nano-oscillator based on a synthetic antiferromagnet free layer which has been studied numerically\cite{solo10}.
Generalizing the procedure of the previous section, the  projection of Eq.\eqref{aka} unto the  two-qubit spin basis gives
 \begin{align}
\hat{\tilde{H}}_{\alpha} &= \sum_{m,n=\pm s}\ket{-m,m}\hat{\tilde{H}}_{mn}\bra{n,-n}\nonumber\\&
=\frac{\Delta}{2}[\hat{\alpha}_x\cos(4s\phi)+\hat{\alpha}_y\sin(4s\phi)]\label{aka5}
\end{align}
In this case the argument of the trigonometric functions can be related to the tunneling of the spins in which $\mathcal{S}_{1,z}$ changes by $2s$ and $\mathcal{S}_{2,z}$ changes by $-2s$.
The complete Hamiltonian of the rotated system must include its mechanical motion about that axis. Thus we have 
\begin{align}
\hat{H}= \frac{\hbar^2(L_{1,z}-L_{2,z})^2}{2I}+\frac{\Delta}{2}[\hat{\alpha}_x\cos(4s\phi)+\hat{\alpha}_y\sin(4s\phi)]
\label{solo3}
\end{align}
where the orbital angular momenta are $L_{i,z}=-i(d/d\phi_i)$, and $I=I_{1,z}-I_{2,z}$ is the relative moment of inertia of the system about the axis of rotation. 
Under a  unitary transformation with the operator $\hat{U}(\phi)=\exp(-2i s\phi\hat{\alpha}_z)$, Eq.\eqref{solo3} becomes
\begin{align}
\hat H\rightarrow \hat{U}^{-1}(\phi)\hat H \hat{U}(\phi)&=\frac{\hbar^2(L_{1,z}-L_{2,z})^2}{2I}+\frac{\Delta}{2}\hat{\alpha}_x\label{rot1}
\end{align}
The first term is as a consequence of mechanical motion of the system which is being rotated.  The total angular momenta  $J_{i,z}= L_{i,z}+\mathcal{S}_{i,z}$ is a conserved quantity. It is crucial to note that in the original problem i.e Eq\eqref{1}, the individual components of the spins $\mathcal{S}_{1,z}$ and $\mathcal{S}_{2,z}$ are not conserved. This leads to an energy splitting $\Delta$, thus allowing for the conservation of the individual total angular momenta $\mathcal{J}_{1,z}$ and $\mathcal{J}_{2,z}$.  If one had included the orbital angular momenta in Eq.\eqref{1}, then the problem is no longer a reduced problem and the total angular momenta $\mathcal{J}_{1,z}$ and $\mathcal{J}_{2,z}$  won't be conserved. It will be interesting to investigate the effect of this inclusion on the energy splitting for large spins. In terms of $J$, Eq.\eqref{rot1} can be written as
\begin{align}
\hat{H}&= \frac{\hbar^2[(J_{1,z}-J_{2,z})^2 + (2s\hat{\alpha}_z)^2]}{2I}+\frac{\Delta}{2}\hat{\alpha}_x\nonumber \\&- {\hbar^24s(J_{1,z}-J_{2,z})\hat{\alpha}_z}/2I
\end{align}

 The simultaneous eigenstate of this system  is given by
\begin{align}
\ket{\Psi_{j_1j_2}}&=\frac{1}{\sqrt{2}}\ket{j_1,j_2}_{l_1,l_2}\otimes(\mathcal{C}_{\uparrow,\downarrow}\ket{\uparrow,\downarrow} +\mathcal{C}_{\downarrow,\uparrow}\ket{\downarrow,\uparrow} )
\end{align}
where $J_{i,z}\ket{j_1,j_2}=j_{i}\ket{j_1,j_2}$.
Diagonalization  of this Hamiltonian yields the corresponding  eigenvalues
\begin{align}
\mathcal{E}_{j_1j_2}=\frac{\Delta}{2}\bigg[\frac{\beta}{2}\lb 1+\frac{\lb j_1-j_2\rb^2}{(2s)^2}\rb\pm \sqrt{1+\frac{\lb j_1-j_2\rb^2}{(2s)^2}\beta^2}\bigg]
\end{align}
with $\beta = 2(2\hbar s)^2/(I\Delta)$. It is noted that the energy levels are degenerate with the sign of $j_1-j_2$ for $j_1,j_2 \neq0$. The coefficients of the wave function are found to be
\begin{align}
\mathcal{C}_{\uparrow,\downarrow}&= \sqrt{1+\beta(j_1-j_2)/\sqrt{(2s)^2+ (\beta(j_1-j_2))^2}}\nonumber\\
\mathcal{C}_{\downarrow,\uparrow}&= \sqrt{1-\beta(j_1-j_2)/\sqrt{(2s)^2+ (\beta(j_1-j_2))^2}}
\end{align}
Now that we have obtained the wave function and its coefficients, the expectation values of the observables $\hat{\alpha}_i$ can be easily evaluated. They are given by
\begin{align}
\braket{\hat{\alpha}_{x}}& = \sqrt{1-\frac{(\beta(j_1-j_2))^2}{(2s)^2+ (\beta(j_1-j_2))^2}}\\
\braket{\hat{\alpha}_{y}}&=0\\
\braket{\hat{\alpha}_{z}}&=  \frac{\beta(j_1-j_2)}{\sqrt{(2s)^2+ (\beta(j_1-j_2))^2}}
\end{align}
\begin{figure}[ht]
\centering
\subfigure[ ]{%
\includegraphics[scale=0.35]{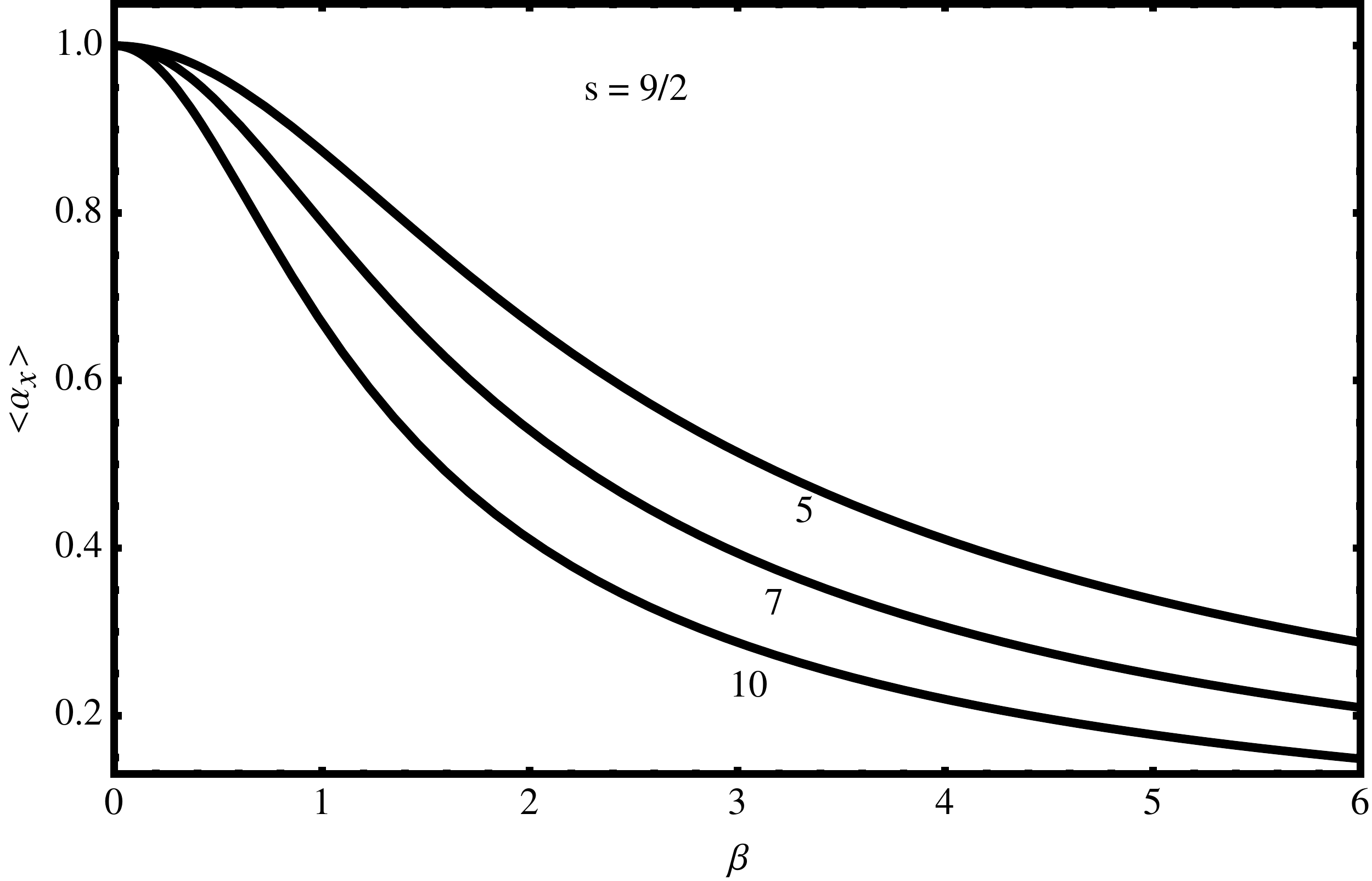}
\label{fr}}
\subfigure[]{%
\includegraphics[scale=0.35]{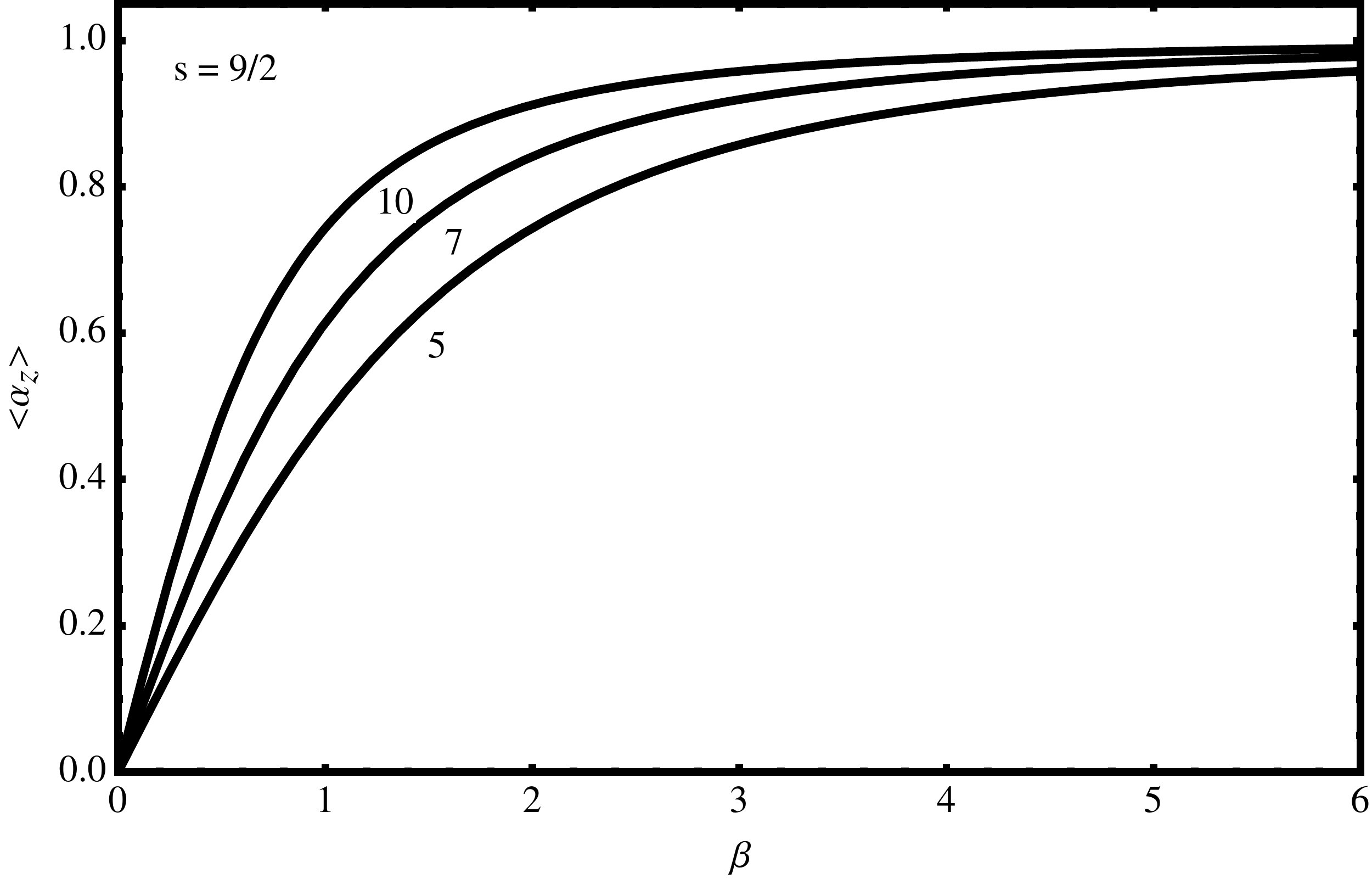}
\label{phase}}
\caption{  The expectation values of $\braket{\hat{\alpha}_{x}}$ and $\braket{\hat{\alpha}_{z}}$ plotted against the parameter $\beta$. Line are labelled with the values of $j_1-j_2$.}
\label{ph}
\end{figure}

In Fig.\eqref{ph} we have shown the plot of the average values of the spins $\braket{\hat{\alpha}_{x}}$ and $\braket{\hat{\alpha}_{z}}$ as a function of the parameter $\beta$. The average value $\braket{\hat{\alpha}_{x}}$ decays with increasing values of $j_1-j_2$, only becomes zero at sufficiently large values of $j_1-j_2$. Notice a similar trend between the coherence factor in Fig.\eqref{coh} and the average $\braket{\hat{\alpha}_{x}}$.  

In the absence of an external magnetic field, the two spins are aligned in an equal and opposite directions so that the total magnetization vanishes, however the staggered magnetic moment does not vanish, which can be computed as
\begin{align}
\mu_s&=g\mu_B\sum_i (-1)^i\braket{\mathcal{S}_{i,z}}
=-\frac{2s\beta g\mu_B(j_1-j_2)}{\sqrt{(2s)^2+ (\beta(j_1-j_2))^2}}
\end{align}
where $g$ is the electrons $g$-factor and $\mu_B$ is the Bohr magneton. In terms of the rotation angle $\phi_i$, Eq.\eqref{rot1} can be written as
\bea
\hat H=\frac{2\hbar^{2}}{I}\frac{d^{2}}{d\phi^{2}}+\frac{\Delta}{2}\hat{\alpha}_x
\eea
This form of the Hamiltonian allows one to solve for the wave function in terms of $\phi$.  By applying the unitary transformation in Eq.\eqref{unitary}, the wave function can be written as
\bea
\Phi^{\pm}(\phi)= Ae^{im_l\phi}+Be^{-im_l\phi}, \quad   \mathcal{E}_{m_l}^{\pm}=\frac{2m_l^2\hbar^2}{I}\mp\frac{\Delta}{2}
\eea
with the boundary condition $\Phi(\phi+2\pi)=\Phi(\phi)$, where $m_l= 0,\pm1,\pm2\cdots$, is the relative quantum numbers. 
 Let us briefly address the question we raised in the introduction. How does a dissipative environment  couple to a rotating molecular dimeric nanomagnet? We reiterate the fact that this question is yet an unsolved problem, however, since the rotation about the easy-axis leaves this axis unchanged, a straight forward generalization of the previous analysis gives
 \begin{align}
\hat{H}&= \frac{\hbar^2[(J_{1,z}-J_{2,z})^2 + (2s\hat{\alpha}_z)^2]}{2I}+\frac{\Delta}{2}\hat{\alpha}_x\nonumber\\&+ \sum_{k}\epsilon_k b_k b_{k}^{\dagger}- {\hbar^24s(J_{1,z}-J_{2,z})\hat{\alpha}_z}/2I
 +\frac{\hat{\alpha}_z}{2}\sum_{k}\gamma_k (b_k+b_{k}^{\dagger})
\label{aka4}
\end{align}
This system involves the interaction of the total angular momenta with the spins, and the spins with the environment. The first step in solving this problem is to find an equivalent  density matrix operator of Eq.\eqref{aka6} from which other observables can be calculated. This analysis can be done in principle.

 {\it{ Conclusions}-}
In conclusion, we have investigated an antiferromagnetically exchange-coupled  dimer of single molecule magnet which possesses a large spin tunneling. Perturbation theory to $2s^{\text{th}}$ order transforms the system into an effective two-state system with the ground and the first excited states being an entangled state of the degenerate eigenstates of the free Hamiltonian with an energy splitting between them. The nature of this Hamiltonian allows us to map the system onto an entangled pseudospin 1/2 two-state system. For an antiferromagnetically exchange-coupled  dimer which is free to rotate about the staggered easy-axis, we obtained the eigenstate and eigenvalues of this system. The average values of the system observables were calculated and plotted with the parameter of the system. Finally, we briefly discussed the environmental influence on a rotating exchange-coupled  dimer. These results can be applied to a free magnetic dimer clusters in a cavity. It is also useful in quantum computation using entangled two-qubit states.

{\it{ Acknowledgments}-}
The author would like to thank NSERC of Canada for financial support with the grant no. R0003315. Also the author acknowledges useful discussions with Joachim Nsofini and Manu Paranjape  .

\end{document}